\documentclass[fleqn,twoside]{article}
\usepackage{espcrc2}

\usepackage{graphicx}
\usepackage[figuresright]{rotating}
\usepackage{amsmath}

\def\gsim{\raise0.3ex\hbox{$>$\kern-0.75em\raise-1.1ex\hbox{$\sim$}}}
\def\lsim{\raise0.3ex\hbox{$<$\kern-0.75em\raise-1.1ex\hbox{$\sim$}}}

\title{ Lorentz symmetry violation, dark matter and dark energy }

\author{Luis Gonzalez-Mestres\address{LAPP, Universit\'e de Savoie, CNRS/IN2P3, B.P. 110, 74941 Annecy-le-Vieux Cedex, France}}

\begin{document}

\begin{abstract}
Taking into account the experimental results of the HiRes and AUGER collaborations, the present status of bounds on Lorentz symmetry violation (LSV) patterns is discussed. Although significant constraints will emerge, a wide range of models and values of parameters will still be left open. Cosmological implications of allowed LSV patterns are discussed focusing on the origin of our Universe, the cosmological constant, dark matter and dark energy. Superbradyons (superluminal preons) may be the actual constituents of vacuum and of standard particles, and form equally a cosmological sea leading to new forms of dark matter and dark energy.
\end{abstract}

\maketitle

\section{Patterns of Lorentz symmetry violation}

A formulation of Planck-scale Lorentz symmetry violation (LSV) testable in ultra-high energy cosmic-ray (UHECR) experiments was proposed in \cite{gon97a,gon97b}. It involves two basic ingredients: i) the existence of a privileged local reference frame (the vacuum rest frame, VRF) ; ii) an energy-dependent parameter driving LSV and possibly making it observable in the ultra-high energy (UHE) region. Then, standard special relativity can remain a low-energy limit in the VRF, contrary to approaches where the critical speed in vacuum is not the same for all standard particles. A simple LSV pattern of the type proposed in \cite{gon97a,gon97b} is quadratically deformed relativistic kinematics (QDRK), where the effective LSV parameter varies quadratically with energy. In the VRF, we can write:

\begin{equation}
E~=~~(2\pi )^{-1}~h~c~a^{-1}~e~(k~a)
\end{equation}
$E$ being the particle energy, $h$ the Planck constant, $c$ the speed of light, $k$ the wave vector, $a$ the fundamental length and $[e~(k~a)]^2$ a convex function of $(k~a)^2$. $a$ can correspond to the Planck scale or to a smaller length scale. Expanding (1) for $k~a~\ll ~1$ , we get \cite{gon97b}:
\begin{equation}
e~(k~a) ~ \simeq ~ [(k~a)^2~-~\alpha ~(k~a)^4~+~(2\pi ~a)^2~h^{-2}~m^2~c^2]^{1/2}
\end{equation}
where $p$ is the particle momentum, $\alpha $ a positive model-dependent constant and {\it m} the mass of the particle. For $p~\gg ~mc$ , one has:
\begin{equation}
E ~ \simeq ~ p~c~+~m^2~c^3~(2~p)^{-1}~-~p~c~\alpha ~(k~a)^2/2
\end{equation}
Kinematic balances are altered, potentially leading to observable phenomena, above a transition energy $E_{trans}$ where the deformation term $-~p~c~\alpha ~(k~a)^2/2$ becomes of the same order as the mass term $ m^2~c^3~(2~p)^{-1}$. For this comparison to make sense, the existence on an absolute local rest frame is a fundamental requirement, even if the ansatz (1)-(3) can be a limit of many different basic theories. 

Assuming exact energy and momentum conservation, two important implications of QDRK for UHE particles were already emphasized in \cite{gon97a} : i) QDRK can lead to a suppression of the Greisen-Zatsepin-Kuzmin (GZK) cutoff \cite{GZK1,GZK2} ; ii) unstable particles live longer at UHE than in standard special relativity, and some of them can even become stable at these energies. For such phenomenological applications, the Earth is assumed to move slowly with respect to the VRF.

Subsequent papers \cite{gon97b,gon97d,gon97cgona} further discussed these issues and that of the universality of the $\alpha $ parameter. Particles with negative values of $\alpha $ could not be stable at UHE, or even at lower energies. Assuming $\alpha $ to be positive, particles with lower values of $\alpha $ would decay into those with larger $\alpha $. There would be at least one stable UHE particle, with the highest value of $\alpha $. A recent paper by Mattingly et al. \cite{Mattingly2009} studies a particular application of this discussion to neutrinos. More obvious possibilities can be considered \cite{gon97b,gon97d,gon97cgona}, taking different families of particles. Spontaneous decays of protons and nuclei by photon emission due to LSV can fake the GZK cutoff \cite{Gonzalez-Mestres2009}. Similarly, spontaneous decays of UHE photons into $e^+~ e^-$ pairs, or the converse effect, may occur at UHE with a moderate difference in $\alpha $ between electrons and photons \cite{gon97d,Gonzalez-Mestres2009}.    

The model previously proposed by Kirzhnits and Chechin \cite{Kir} does not lead to such predictions and is unable to produce the suppression of the GZK cutoff \cite{gon02}. The situation is similar for standard Doubly Special Relativity (SDSR) patterns \cite{SDSR}. In both cases, the laws of Physics are assumed to be the same in all inertial frames. Thus, a symmetry transformation can turn the UHE particle into a less energetic one, leading to a situation where LSV is weaker. We call Weak Doubly Special Relativity (WDSR) our approach that, instead, assumes the existence of a local privileged VRF where special relativity is a low-energy limit. In this case, the laws of Physics are not exactly identical in all inertial frames. The QDRK discussed here is a particular form of WDSR.

In 1971, Sato and Tati \cite{Sato} suggested that the possible absence of the GZK cutoff be explained by an {\it ad hoc} suppression of hadron production in UHE collisions related to a cutoff in the Lorentz factor below $\approx 10^{11}$. Pion production would then be precluded above $\approx 10^{19}$ eV. Our proposal does not involve such a hypothesis. UHE protons can release a substantial part of their energy in the form of pions when colliding with a photon, provided the photon energy is larger than both the proton mass term and the LSV deformation term of proton kinematics. In QDRK, contrary to the Sato-Tati scenario, the possibility to suppress the GZK cutoff is linked to the exceptionally low energy of cosmic microwave background (CMB) photons and not to an intrinsic cutoff on UHE hadron production.

From a cosmological point of view, it seems reasonable in WDSR patterns to associate to Planck time, or to a smaller time scale, the arisal of LSV in the structure of the physical vacuum. This time scale can be $\approx ~a ~c^{-1}$, or $\approx ~a ~c_s^{-1}$ if a critical superluminal speed $c_s$ exists as in the superbradyon hypothesis. In all cases, the internal structure of the physical vacuum and the history of the Universe may contain strong remnants of their early, Lorentz violating, formation that are not incorporated in standard particle physics and cosmology. Such remnants can also exist as free particles in the present Universe or have influenced structure formation. Issues such as the cosmological constant, inflation, dark matter and dark energy may crucially depend on the role of this new physics.

\section{Superbradyons}

Standard preon models \cite{preons} assumed that preons feel the same minkowskian space-time as quarks, leptons and gauge bosons, and carry the same kind of charges and quantum numbers. But there is no fundamental reason for this assumption.

Superbradyons \cite{gon97a,Gonzalez-Mestres2009,gonSL2} would have a critical speed in vacuum $c_s ~\gg ~c$ possibly corresponding to a superluminal Lorentz invariance (a symmetry of the Lorentz type with $c_s$ instead of $c$). They may be the ultimate constituents of matter, generate LSV for "ordinary" particles (those with critical speed equal to $c$) and even obey a new mechanics different from standard quantum mechanics \cite{gon09b,gon09c}. 

After Planck time, superbradyons and "ordinary" particles may coexist in our Universe. In this case, contrary to tachyons, superbradyons would have positive mass and energy, explicitly violate standard Lorentz invariance and be able to spontaneously emit "Cherenkov" radiation in vacuum in the form of "ordinary" particles. Single superbradyons would in general have very weak direct couplings to the conventional interactions of standard particles. In particular, they would not obey the standard relation between inertial and gravitational mass. Then, the vacuum itself may present a similar behavior leading to important effects. We do not consider here : i) the possible existence of several superbradyonic sectors of matter ; ii) the possible differences between the critical speed of superbradyons in their original dynamics without conventional matter, and the actual superbradyon critical speed in our Universe. 

In the VRF, the energy $E$ and momentum $p$ of a free superbradyon with inertial mass $m$ and speed $v$ would be : 
\begin{eqnarray}
E~=~c_s~(p^2~+~m^2 ~c_s^2)^{1/2} \\
p~=~m~v~(1 ~-~v^2~c_s^{-2})^{-1/2}
\end{eqnarray}

and, for $v~\ll ~c_s$ , we get :
\begin{eqnarray}
E~\simeq ~m~c_s^2~+~m~v^2/2 \\
p~\simeq ~m~v
\end{eqnarray}

The kinetic energy $E_{kin}$ is then $E_{kin}~\simeq ~m~v^2/2$, and  $E_{kin}~\gg ~p~c$ for $v ~\gg ~c$. Superbradyons with $v ~> ~c$ are kinematically allowed to decay by emitting standard particles. As lifetimes for such processes can be very long because of the weak couplings expected, the decays may still exist in the present Universe and play a cosmological role. A superbradyon decay would emit a set of conventional particles with total momentum $p_T~\ll ~E_T ~c^{-1}$ where $E_T$ is the total energy of the emitted particles. Such an event may fake the decay of a conventional heavy particle or the annihilation of two heavy particles, or contain pairs of heavy particles of all kinds. The situation would be similar if the superbradyon decays into one or several lighter superbradyons plus a set of conventional particles, or if two superbradyons annihilate or interact emitting "ordinary" particles. 

Superbradyons can thus provide an unconventional form of dark matter in our Universe, and even produce observable signatures at comparatively low energies. Those with $v ~\simeq ~c$ would form a stable sea, except for annihilations. The possibility that superbradyons be a source of standard UHECR was also considered in \cite{gon96}.

\section{Experimental considerations}

As emphasized in \cite{gon97b,gon97d,gon97cgona}, a LSV $\approx ~10^{-6}$ at the Planck scale in QDRK for the highest-energy particles would be enough to suppress the GZK cutoff. Thus, such an approach to LSV is currently being tested by ultra-high energy cosmic-ray (UHECR) experiments. Data and analyses from the AUGER \cite{AUGER} and HiRes \cite{HiRes} collaborations possibly confirm the existence of the GZK cutoff. Significant bounds on LSV scenarios will emerge from these data. However, a large domain of LSV patterns and values of parameters will still remain allowed, even for QDRK models \cite{Gonzalez-Mestres2009,Gonzalez-Mestres2008}. 

A crucial issue is that of the composition of the UHECR spectrum. The AUGER Collaboration has recently reported \cite{Ulrich} a systematic inconsistency of available hadronic interaction models when attempting to simultaneously describe the observations of the $X_{max}$ parameter and the number $N_\mu $ of produced muons. Data on $X_{max}$ suggest UHECR masses to lie in the range between proton and iron, while $N_\mu $ data hint to heavier nuclei. 

If the highest-energy particles are nuclei, the AUGER and HiRes results can still be compatible with a LSV $\approx ~1$ at the Planck scale ($\alpha ~a^2 ~ \approx ~a_{Pl} ^2$, where $a_{Pl}$ is the Planck length) for quarks and gluons. But even assuming that a significant part of the highest-energy particles are protons, the actual bounds on LSV for quarks and gluons will depend on the internal structure of the proton at UHE. Possible spontaneous decays of UHECR protons and nuclei, but also of photons, must equally be considered \cite{Gonzalez-Mestres2009}. Further explorations will thus be required, including satellite experiments \cite{gonSL3}. 

Other LSV patterns (e.g. LDRK, linearly deformed relativistic kinematics, where the parameter driving LSV varies linearly with energy) were discarded in \cite{gon97a} and \cite{gon97b,gon97d,gon97cgona}, as they lead to too strong effects at low energy if the parameters are chosen to produce observable effects at UHECR energies. Hybrid models with high-energy thresholds can still be considered \cite{gonSL3}, but will not be dealt with here. Inhibition of synchrotron radiation in LSV patterns was predicted and studied in our 1997-2001 papers for QDRK at UHE \cite{gon97b,gon97d,gon97cgona}, leading also to tests like that presented in \cite{Jacobson} for 100 MeV synchrotron radiation from the Crab nebula with the same kind of calculation in a version of LDRK.

The energy balances used here involve energies that are very small as compared to those of the particle interactions considered. Therefore, UHECR experiments can also be viewed as tests of energy and momentum conservation and of the validity of quantum mechanics at UHE \cite{gon09b,gon09c}. These phenomenological aspects deserve further study.

As an alternative to standard dark matter, cosmic superbradyons can potentially provide \cite{gon09b} an explanation to the electron and positron abundances reported by PAMELA \cite{PAMELA}, ATIC \cite{ATIC}, Fermi LAT \cite{FERMI}, HESS \cite{HESS} and PPB-BETS \cite{PPB-BETS}. A cosmological sea of superbradyons would still be decaying through the emission of "Cherenkov" radiation in vacuum or releasing conventional particles for some other reason. Whether or not data on electrons do exhibit a bump between 300 GeV and 600 GeV \cite{Malyshev} does not change this conclusion. To date, the interpretation of such experimental results in terms of standard dark matter remains unclear \cite{deBoer}. More conventional astrophysical interpretations of these data have been considered in \cite{FERMI,altern}.

The possible experimental consequences of a superbradyon era in the early Universe deserve further investigation \cite{gon09b,gon09c}, as well as the role of superbradyons in the present vacuum, its connection to dark energy effects and the possibility that superbradyonic matter replaces some of the scalar field condensates of standard physics and cosmology. 
 
\section{Post Scriptum after publication in AIP Conference Proceedings}

The version v1 of this paper \cite{Gonzalez-Mestresv1} (December 2009), that we have reproduced here, corresponds to the text published in the Proceedings of the 2009 Invisible Universe Conference \cite{Gonzalez-MestresProc}. In what follows, we add some comments and references on more recent information and new ideas.

Recent LHC data on the quark-gluon plasma \cite{LHC}, confirming previous RHIC results at lower energies \cite{RHIC}, seem to suggest that some aspects of standard cosmology may have to be reconsidered. In this respect, it seems worth further emphasizing the potential role of vacuum in issues concerning cosmology and the structure and behaviour of matter at very small distances \cite{Gonzalez-Mestres2010a,Gonzalez-Mestres2010b}. This is particularly relevant for tests of the validity of fundamental laws (relativity, quantum mechanics, energy and momentum conservation...) but it also applies generally to other basic questions in particle physics and astrophysics. To date, essential issues remain unsettled concerning the vacuum structure, dynamics and evolution between the beginning of our Universe and the present epoch.

Thus, caution is required when inferring consequences from LHC and other accelerator results for cosmological eras (f.i. concerning the possible properties of a quark-gluon plasma), as the present physical vacuum is not the same and we are living in a much colder Universe. 

Energy thresholds in LSV were considered in \cite{gonSL3}, but a more radical version has been recently suggested by Anchordoqui et al. \cite{Landsberg} where the number of effective space dimensions decreases with the energy scale through scale thresholds. This new LSV pattern may be a candidate to explain the jet alignement possibly observed by Pamir and other experiments above $\sim ~ 10^{16}$ eV \cite{mountain,stratosph}. Actually, as shown in \cite{Gonzalez-Mestres2010a}, missing transverse energy in cosmic-ray interaction jets above some energy scale ($\sim ~ 10^{16}$ eV to fit Pamir data) can be due to the production of superbradyonic objects (waves, particles...) involving a small portion of the total energy and a negligible fraction of momentum. Such a phenomenon, that can be related to energy capture by vacuum, would naturally lead to elongated jets. Subsequent polarization effects inside vacuum and secondaries can then contribute to planar jet alignment \cite{Gonzalez-Mestres2010b}. 

Similar phenomena can also be present at higher energies than those considered by the analyses of the Pamir data. Furthermore, at a scale above the highest observed cosmic-ray energies, a fall of the cross-sections of cosmic rays with the atmosphere is also expected \cite{gon97d}. 

As previously foreseen for LSV with QDRK \cite{gonSL2,gonSL3} considering the role of the deformation and the new approach to vacuum structure, the recent suggestion by Anchordoqui et al. is also claimed to potentially cure ultraviolet divergences in field theory \cite{Landsberg} or solve the cosmological constant problem \cite{GarciaA}. More generally, a new approach to quantum field theory seems necessary at the present stage for very high energy interactions and vacuum structure. 

As in other periods in the development of particle physics, condensed matter physics can be a useful guide to elaborate new concepts adapted to current open questions \cite{gon97a,gonSL2,Gonzalez-Mestres2010b}. 

Completing \cite{Gonzalez-Mestres2009}, reference \cite{Gonzalez-Mestres2010b} also updates our analysis of the present situation concerning data from UHECR experiments and the tests of special relativity in the GZK energy region.

Another recent development is the publication of the 7-year WMAP data and analyses \cite{WMAP}, leading in particular to the suggestion by Gurzadyan and Penrose \cite{Gurzadyan1} that concentric circles in WMAP data exhibiting anomalously low temperature variance may provide evidence of violent pre-Big-Bang activity. For further discussion on the WMAP concentric circles, see \cite{Moss,Wehus} and \cite{Gurzadyan2,Gurzadyan3}. In an approach based on conformal cyclic cosmology \cite{Penrose1,Penrose2}, Gurzadyan and Penrose consider black-hole encounters in a previous aeon, possibly producing such signals and questioning the validity of standard inflation. Noncyclic cosmologies based on superbradyonic dynamics and a superbradyon era replacing or preceding the standard Big Bang have also been proposed as an alternative to inflation \cite{gonSL2,gon09b,gon09c,Gonzalez-Mestres2010b} and can naturally provide explanations to this kind of observations. Then, the possible WMAP signals may have been generated in the previous superbradyonic universe or during the transition from superbradyonic to ordinary matter. More generally, LSV patterns with a fundamental length smaller than the Planck length can also produce similar phenomena in the transition to Planck scale. 

Furthermore, the discussion on the WMAP circles raises more globally the question of the possible lack of randomness of WMAP data. According to Gurzadyan et al. \cite{Gurzadyan2}, the cosmological sky is a weakly random one where "the random perturbation is a minor component of mostly regular signal". Such a situation, if confirmed, may open unprecedented ways for new cosmological phenomenology, including tests of all kinds of alternatives (string-like, LSV with a new fundamental length, superbradyons...) to the standard Big Bang and inflationary scenarios.

Superbradyon patterns and LSV generated beyond Planck scale can thus provide coherent, non cyclic, alternatives to current cyclic cosmologies describing pre-Big-Bang scenarios. Strings would then naturally become composite objects or be replaced by new theoretical concepts at the Planck scale or beyond. 

Although the superbradyon hypothesis and QDRK do not by now provide a theory of matter, they should not be considered as a purely phenomenological framework. Superbradyons imply a new conceptual approach to preons and to the structure of vacuum and conventional particles. QDRK is the natural choice for the deformed relativistic kinematics if the particles of the standard model are assumed to be composite \cite{gon97a,Gonzalez-Mestres2010a,Gonzalez-Mestres2010b}. It remains to be determined to what extent superbradyons can be represented as particles, if they indeed obey quantum mechanics and a new Lorentz symmetry with $c_s$ as the critical speed, if or how energy and momentum are conserved at UHE scales... The study of UHE physics and of the possible superbradyonic domain will require a long term experimental and phenomenological work involving UHECR experiments and possibly the study of cosmic microwave background radiation. LHC experiments and dark matter studies can also be relevant in specific cases \cite{gonSL2,Gonzalez-Mestres2010b,gon97e}.   

A possible solution to the dark energy problem may also involve geometric patterns of space-time structure like that considered in \cite{gon97c}, where it was suggested to replace the standard real four-dimensional space-time by a SU(2) spinorial one. Then, spin-1/2 particles would be representations of the actual group of space-time transformations. From a space-time spinor $\xi $, the positive scalar $\mid \xi \mid ^2$ $=$ $\xi ^\dagger \xi $ can be obtained where the dagger stands for hermitic conjugate. In a simple physical interpretation of the geometry at cosmological scale, a positive cosmic time $t~=~\mid \xi \mid$, the spinor modulus, can be defined which leads in particular to a naturally expanding Universe. To define local space coordinates in this approach, one can consider a spinor $\xi _0$ (the observer position) on the $\mid \xi \mid $ = $t_0$ hypersphere. Writing, for a point $\xi $ of the same spatial hypersphere :
\begin{equation}
\xi ~=~ U\xi _0
\end{equation}
where $U$ is a SU(2) transformation :
\begin{equation}
U~=~exp~(i/2~~t_0^{-1}~{\vec \sigma }.{\vec {\mathbf x}})~
\equiv U({\vec {\mathbf x}}) 
\end{equation}
and ${\vec \sigma }$ the vector formed by the Pauli matrices, the vector ${\vec {\mathbf x}}$, with $0~\leq x$ (modulus of ${\vec {\mathbf x}}$)  $\leq$  $2\pi t_0$, can be interpreted as the spatial position vector at constant time $t_0$. A $2\pi $ rotation ($x~=~2\pi t_0~,~U~=~-1$) changes the signs of $\xi _0$ and $\xi $ simultaneously, and leaves ${\vec {\mathbf x}}$ invariant. Conventional local space coordinates are obtained for $x ~\ll ~t_0$. More details can be found in \cite{gon97c}. 

Then, the local time scale and physical laws will be determined by physical processes and vacuum structure in the space-time region close to $\xi _0$. The very small $t$ region, associated in principle to the Big Bang, may be deformed or altered by superbradyon dynamics and by LSV generated beyond Planck scale.

Reformulating general relativity, quantum field theory, possible LSV and cosmology in such spinorial coordinates can be a useful exercise, and will be further dicussed elsewhere.

\section{Note added to the Post Scriptum}

Further discussion on possible signatures of cyclic cosmologies and on the degree of randomness of WMAP data can be found in \cite{Gurzadyan4,Gurzadyan5}, \cite{Eriksen,Naess} and \cite{Kocharyan,Staig}.

In page 2 of this paper, the sentence : "In 1971, Sato and Tati \cite{Sato} suggested that the GZK be explained by an {\it ad hoc} suppression of hadron production in UHE collisions" must be understood as referring to an explanation of the possible absence of the GZK cutoff suggested in \cite{Sato}. According to Sato and Tati, the GZK cutoff would possibly be absent due to an impossibility to produce pions above $\approx 10^{19}$ eV. As stressed above, such a constraint does not exist in QDRK. 

\subsection{Are "fundamental principles" fundamental ?}

In connection with such a debate, it seems worth emphasizing again that properties of the standard particles and of our Universe nowadays considered as fundamental principles of Physics may actually be of composite origin or the natural result of a primordial evolution. 

As an example, Lorentz symmetry (even approximate or as a low-energy limit) can be a natural outcome of many primordial scenarios due to the stability of its kinematical structure as compared to other space-time geometries (f.i. euclidean) that would make matter unstable through spontaneous particle production and are also disfavoured by phase space considerations.

A simple illustration of metrics instability can be provided by a kinematics of the form:
\begin{equation}
E^2 ~+~ (p~c')^2 ~=~ M^2~c'^4
\end{equation}
\noindent where $c'$ is a characteristic speed and $M$ a mass. 

With such a kinematics, the particle phase space is strongly restricted as compared to a minkowskian metrics. Therefore, possible metrics fluctuations at the Big Bang scale or at some previous fundamental scale can eventually favour the transition to a space-time structure of the standard Lorentz type. Furthermore, if the kinematics defined by (10) applies, the vacuum can in principle spontaneously emit particles with momentum $p ~=~ M~c'$ and is therefore naturally unstable. Again, the dynamical transition to a minkowskian metrics appears as a logical phenomenon.

Such considerations apply not only to conventional particles, but also to superbradyons. They may naturally lead to the patterns considered in our papers since 1995 \cite{gonSL2,gonSL3} where several sectors of matter may exist with different critical speeds in vacuum but possibly obeying in all cases symmetries of the Lorentz type in the low-momentum limit. 

Then, it is possible that effective space-time configurations of a different form can only exist temporarily in specific transition situations (the nucleation of our Universe in superbradyonic matter ?).  

\subsection{Our Universe and pre-Big Bang superbradyonic matter}

As emphasized in version 5 of \cite{gon09b}, in a composite pattern where the standard particles and our Universe would be generated from a superbradyonic medium, the standard gauge bosons do not need to be associated to local symmetries in the conventional sense. 

Superbradyons would not be mere building blocks in the usual sense of conventional preon \cite{preons} patterns. They can instead form a real new phase of matter where conventional particles would be collective excitations of a large medium. This may have strong implications for standard gauge theories. 

Then, the Big Bang may correspond to the nucleation of a new phase of matter (the "conventional" one as seen by human observers) in a superbradyonic pre-Universe. The new phase subsequently expands in superbradyonic matter. Several scenarios can then be imagined, such as for instance :

i) The Universe just expands as a nucleated phase, the nucleation corresponding basically to the transformation of the superbradyonic ground state into our physical vacuum. Then, if the expansion of the conventional vacuum occurs in a locally random way, the dilation of the apparent space can be a natural consequence leading to Lema{\^i}tre's relation between distances, velocities and redshifts \cite{Lemaitre1927}. The dark energy effect can be related to the energetic balance of the phase transition between the superbradyonic ground state and the conventional vacuum.

ii) Our three-dimensional Universe is similar to a new kind of three-dimensional soliton in a space with four or more dimensions. Soliton - antisoliton pair production may then account for matter-antimatter separation and CP violation. Such solitons would have an internal structure, initially very hot and with a very small spatial size, and expand in the superbadyonic matter. Again, Lema{\^i}tre's law can result from relaxation through the expansion of the conventional physical vacuum from the phase transition of the superbradyonic matter.

iii) Our Universe is generated within a spinorial space-time structure like that described in \cite{gon97c}, also considered in \cite{Gonzalez-Mestres2010b} and in the above Post Scriptum. Then, the standard space coordinates would be defined locally and Lema{\^i}tre's law would correspond to conventional matter following basically straight lines along the orthogonal (cosmic time) direction. The relation between local and cosmic time would depend on the dynamics of the Universe expansion.

Scenario iii) appears particularly well suited to describe half-integer spin as a real angular momentum and formulate a primordial version of quantum mechanics. Incorporating a similar geometry into scenarios i) and ii) is also feasible. 

In i), a physical privileged point or small zone inside our Universe may correspond to its origin. In ii), a similar point or small zone can exist in the superbradyonic matter.  Matter and energy exchanges between the conventional Universe and outside superbradyonic matter cannot be excluded. 

\subsection{Are gauge field theories really exact theories ?}

Standard quantum field theory is based on local gauge symmetry, where the existence of gauge bosons is required by invariance under local symmetry transformations. But does this concept apply to Physics at Planck scale or beyond it ? How should it be possibly modified ? And how would possible new Physics extrapolate to lower energy scales, especially if the string model is not the ultimate theory \cite{Gonzalez-Mestres2010b} ?

In an underlying superbradyonic picture, taking for simplicity a lattice description, our conventional gauge interactions can at the origin be associated to nearest-neighbour couplings (local potentials) describing the interaction between different local superbradyonic excitation modes. At such stage and scales, quantum mechanics would not necessarily apply and virtual standard particles are not yet needed. 

Then, it is perfectly conceivable that a new dynamics be at work where the standard vector boson fields carrying the conventional gauge forces would be generated from the superbradyonic matter and degrees of freedom only in specific situations. Basicallly, when these nearest-neighbour couplings turn out to depend on position, time and direction due to the material existence of propagating vacuum excitations (the standard particles) involving the same family of local excitation modes. In the absence of surrounding conventional particles, the vacuum structure would be different from that suggested by standard quantum field theory.

In this case, for instance, the Higgs bosons would not need to be statically materialized as a permanent condensate in the physical vacuum of our Universe. The Higgs mechanism can instead be part of a dynamical reaction of vacuum to the physical presence of gauge bosons and of other standard particles, and occur at the set of frequencies required by the interaction between the standard particles involved. Otherwise, the original superbradyonic matter and degrees of freedom can fill the physical vacuum in a different way. Similar considerations would apply to the zero modes of standard bosonic oscillators. Contrary to conventional quantum field theory, only the wavelengths and frequencies excited by the presence of standard matter would contribute to the conventional local vacuum condensate. 

Such a scenario would imply that the energy difference between the superbradyonic ground state and our vacuum is small, and that the superbradyonic structure does not couple to gravity and to other conventional interactions in the same way standard matter (including the vacuum) does. It does not contradict experimental results \cite{Lamoreaux,Wilson} on the Casimir effect \cite{Casimir}. It naturally generates new and simpler solutions to the cosmological constant problem, as well as to the dark energy and dark matter issues, and leads to a new and safer approach to the question of ultraviolet divergencies.

\subsection{UHECR tests of fundamental principles}

Two recent papers by the AUGER collaboration are \cite{AUGER2011a,AUGER2011b}. See also \cite{Lemoine,Giacinti}. The crucial question of UHECR cosmic-ray composition remains by now to be settled.

Saveliev, Maccione and Sigl \cite{Saveliev} have recently used for phenomenological purposes the QDRK model we suggested in 1997 \cite{gon97a,gon97b} and the subsequent pattern for nuclei we worked out \cite{gon97f} where the effective LSV parameter varies basically like $\sim N^{-2}$, $N$ being the number of nucleons (see also \cite{Gonzalez-Mestres2009}). They also use the vacuum Cherenkov effect from superluminal particles already considered in \cite{gonSL95,gonSL96} and more recently in \cite{Gonzalez-Mestres2009,Gonzalez-Mestres2010b}. For positive values of $\alpha $, Saveliev et al. they get upper bounds similar to those suggested in \cite{Gonzalez-Mestres2009}. Concerning spontaneous decays in DRK patterns, see also \cite{gon97g} and \cite{gon05-06}. 

However, Saveliev et al. also consider possible negative values of $\alpha $ and one might infer from their result that, if $^{56}$Fe is present in the UHECR spectrum at $E~=~10^{20}$ eV, then $- ~\alpha $ can be as large as $\simeq ~4$ for single nucleons taking $a$ to be the Planck length $a_{Pl}$. 

In this respect, it must be noticed that for $\alpha $ = - 4, one has a positive deformation term $\Delta E ~= ~-~p~c~\alpha ~(k~a)^2/2~\simeq ~ 13$ eV for $E$ (nucleon) = $10^{19}$ eV, whereas the nucleon mass term at this energy is $m^2~(2~p)^{-1}~ ~\simeq ~ 4$x$10^{-2}$ eV. With such a kinematics, a $10^{19}$ eV proton can spontaneously decay, for instance, into two protons plus an antiproton. Other spontaneous decays will also occur, with processes depending on the value of $\alpha $ for pions, protons, electrons... Massless photons with a positive value of $- ~\alpha $ would be unstable at all energies due to spontaneous decays. 

As the photon cannot in principle have a negative value of $\alpha $, we expect all charged particles with a negative $\alpha $ to spontaneously decay by emitting a photon above some energy. Such a decay would be kinematically allowed for protons above $\simeq $ 4x10$^{19}$ eV if $\alpha $ (photon) = 0 and - $\alpha $ (proton) $\simeq $10$^{-4}$ which corresponds to the lowest positive bound on - $\alpha $ considered by Saveliev et al. As photons with any energy can be spontaneously emitted in this case, we expect such decays to preclude the astrophysical propagation of charged particles at energies where the positive deformation term is larger than the mass term. 

Furthermore, a negative $\alpha $ would make more difficult particle acceleration at astrophysical sources and increase synchrotron radiation by the converse effect to that described in \cite{gon00Sync}.  

As further emphasized in \cite{Gonzalez-Mestres2010b}, relativity is not the only fundamental principle of Physics that can be tested by UHECR experiments. If a privileged inertial local rest frame exists, as systematically assumed in our papers, all basic principles of our current understanding of Nature may undergo changes or transitions at the Planck scale or at some other scale, possibly leading to observable signatures at very high energy.   

If quantum mechanics has been generated in our Universe at the Planck scale or somewhere between a "nucleation time" and the Planck time, the standard quantum mechanics is expected to be deformed as considered above and in \cite{gon09c,Gonzalez-Mestres2010b}. A more complete analysis of possible patterns of quantum mechanics deformation in cosmological scenarios with a VRF is required at that stage in view of test with UHECR.

Similarly, if as suggested above the Higgs boson and the standard bosonic zero modes are not permanently condensed in vacuum, and if the energy difference between the superbradyonic ground state and our vacuum is actually small, nonstandard fluctuations of the vacuum structure may occur in large zones of our Universe and be felt by conventional particles during intergalactic or galactic propagation. Again, UHECR experiments may be sensitive to such effects. More details will be presented elsewhere.  

\bibliographystyle{aipprocl} 
 
\end{document}